# Self Organization Agent Oriented Dynamic Resource Allocation on Open Federated Clouds Environment

Kemchi Sofiane
LIRE Labs
University of Constantine 2
25000 Constantine, Algeria
Sofiane.kemchi@univ-constantine2.dz

Abdelhafid Zitouni
LIRE Labs
University of Constantine 2
25000 Constantine, Algeria
Abdelhafid.zitouni@univ-constantine2.dz

Mahieddine Djoudi
TechNE Labs
University of Poitiers
86073 Poitiers Cedex, France
mdjoudi@univ-poitiers.fr

*Abstract*— To ensure uninterrupted services to the cloud clients from federated cloud providers, it is important to guarantee an efficient allocation of the cloud resources to users to improve the rate of client satisfaction and the quality of the service provisions. It is better to get as more computing and storage resources as possible. In cloud domain several Multi Agent Resource Allocation methods have been proposed to implement the problem of dynamic resource allocation. However the problem is still open and many works to do in this field. In cloud computing robustness is important so in this paper we focus on auto-adaptive method to deal with changes of open federated cloud computing environment. Our approach is hybrid, we first adopt an existing organizations optimization approach for self organization in broker agent organization to combine it with already existing Multi Agent Resource Allocation approach on Federated Clouds. We consider an open clouds federation environment which is dynamic and in constant evolution, new cloud operators can join the federation or leave this one. At the same time our approach is multi criterion which can take in account various parameters (i.e. computing load balance of mediator agent, geographical distance (network delay) between costumer and provider...).

*Keywords*— *Open Federated Cloud Computing, Multi Agent Resource Allocation, Adaptive Multi Agent System, Multi Agent Organization.*

I. INTRODUCTION

Cloud computing is evolving rapidly, as a large scale computing paradigm driven by scale economies, cloud computing allows cloud providers to reserve cloud resources based on time-varying needs. Generally cloud users pay for cloud resources (i.e., CPU, storage, and network bandwidth), platforms, and application services in a pay-as-you-go model, which can help enterprises reduce enormous upfront infrastructure investments [1].

A next step in this evolution is to have many providers of Cloud services, We speaks about a federation of Cloud. the federation of cloud resources offers clients in addition of sharing wide range of resources, the opportunity to choose the best cloud services provider not only in terms of cost but also in term of service flexibility and availability to meet a particular business or technological need within their organization [2], [3].

Facing environment's changes in open federated clouds, the method use self organization property in multi-agent system to allocate providers resources. Users, providers and brokers are considered as agents : User (CA : Consumer Agent), Broker (RBA : Resource Brokering Agent) and Provider (RPA : Resource Provider Agent). RBA will assume the complicate resource allocation task between providers in clouds federation and users. This method is generally known as Multi-agent Resource Allocation (MARA), RBA resource allocation strategy must maximize the benefit in resource allocation, take advantage from the large services which the federation can offer to realize the best quality of the service provisions and improve the rate of client satisfaction [2].

In reality business is a very dynamic environment, new cloud providers can enter to the market and offer their new services, while other cloud providers can leave it. If we consider an open federated clouds , the allocation resource method must be robust capable to deal with changes to permit the join of new cloud operators to the federation or departure from this one. The allocation resource method must adapt itself, in case where RBA and after checking all his clouds provider contact's list is always incapable to satisfy the client resource demand and can't find the client request resource (i.e., resource still unavailable, price not suitable, specific provider leaves the federation, entry of new provider with a particular business or technological need which is not still well known by all brokers...). In such situation RBA organization will change their self organization, adaptation mechanism in our approach is done by delegation of a submitted customer request from one RBA, to another RBA, via multi criterion migration of a submitted customer request, it can take in account various parameters (computing load balance of mediator agent and geographical distance (network delay) between costumer and provider...), for this aim we propose run-time algorithm to implement our self organization mechanism. Migration in our approach preserve continuously the organization's coherence, some preventive coherence constraints are verified before request's migration and other corrective coherence constraints are verified after request's migration [2], [4], [5].

The price of the resources in a cloud is generally variable and based on a demand-supply model. In open federated clouds, users request more than one type of resources from different providers which can leave or join federation. So they need continuously an update and fresh information about all current available service providers and the status of each already present provider in the federation or probably newly joining this one. Choosing the best provider is very difficult because they don't know the dynamic price of each resources in different clouds also due to dynamic environment of an open federated clouds, users may miss a specific service from one, leaving or new coming, specific provider and so failing to satisfy a specific customer demand while in cloud computing we aim to increase the rate of client satisfaction and improve the quality of the service provisions [2]. In this project we suggest a Self Organization Agent Oriented Dynamic Resource Allocation on Open Federated Clouds Environment.

## II. RELATED WORKS

Choosing the best provider in federated clouds is a very difficult task because users don't know price of each resources in different clouds, the price is determined dynamically based on a demand-supply model. Haresh M V et. al.[2] suggest a broker based multi-agent system. In this method, the user didn't care about both identity of the cloud provider belonging to a federated clouds and the location of the resources needed. To know which cloud service provider will perform the request isn't much important as the consumer have to get the resources with the minimum price. However as this approach is based on non flexible system model so there is no possibility to the federated clouds to acquire new providers with challenged price in addition services offered are fixed from the start and the federated clouds can't provide new kind of services. Some approaches brings new concepts as 'borrowing' and 'leasing' resources from and to other Clouds. In order to serve extra client requests Yisheng Wang et. al.[6] focuses on borrowing computing resources from foreign Clouds when home Cloud is subject of congestion and 'leasing' resources to foreign Clouds when home Cloud is free. As dependency relationships in cloud federation constitute a potential risk factors to the performance of the system and to deal with, they use some mechanisms which involve dynamic resource arrangement in establishment and deconstruction of a Cloud Federation. In order to include in the federation different cloud middleware Giuseppe Andronico et. al.[4] provide a concrete model that looks at heterogeneous cloud systems. in their work they describe a model able to consider all implications in accomplishing and managing Dynamic Cloud Federations. the model target small cloud operators allowing them to easily join and leave the federation. Jie Xu et. al.[7] propose an approach of self-organizing based on multi-agent systems To achieve the required macroscopic properties of locally interacting agents in cloud market. They suggest three-layered self-organizing multi-agents mechanism to support cloud commerce parallel negotiation activities. Their consumer model running mechanism use an algorithm as a protocol of negotiation. As we see in [2], [4], [6] and [7] most of the works in federated clouds concerns the study of fixed system models while open clouds federation is more faithful to business reality. At present we need a system able to efficiently support the collaboration between different cloud providers focusing on various aspects of the federation like ensuring flexibility and evolution features of clouds federation.

Few works like Fu Hou et. al.[8] take into account openness features in clouds federation. They present a self-management approach for the cloud services, with an autonomous and context-aware management of the resources by employing a number of service agents in the cloud environment. Based on this approach, they present a cloud-oriented services self-management framework with suitable mechanisms for service aggregations and service provisions, two supporting algorithms are designed to implement the proposed services self-organization process and also the service provision process. In [8] several types of agents are involved in the cloud services self-management such as service manager agent, manager center agent, and service broker agent. The service manager consider an optimization and balance method which is insufficient, more and various range of criteria should be take in consideration to cover different sides of the system as : QoS, computing load balance : (workload parameter), geographical distance "network delay" between costumer and provider : (time response).

## III. SELF ORGANIZATION AGENT ORIENTED DYNAMIC RESOURCE ALLOCATION ON OPEN FEDERATED CLOUDS ENVIRONMENT

Cloud Computing attract more and more business users where they can purchase different resources (i.e., CPU, storage, and network bandwidth), platforms, and application services in a pay-as-you-go model, the price can be fixed or variable. However, generally the objective of costumers and enterprises is finding the lowest resource price with highest speed, from one or many cloud providers. So, the federation of cloud resources offers clients in addition of sharing wide range of resources, the opportunity to choose the best cloud services provider not only in terms of cost but also in term of service flexibility and availability to meet a particular business or technological need within their organization.

In [2] Haresh M V et. al. considered the case of federated clouds based on static system model but actually world of business is a highly variable environment. In this paper, we will consider more flexible system model which take in account an open federated clouds to permit movement of cloud providers on federated clouds and tolerate entrance or departure of clouds.

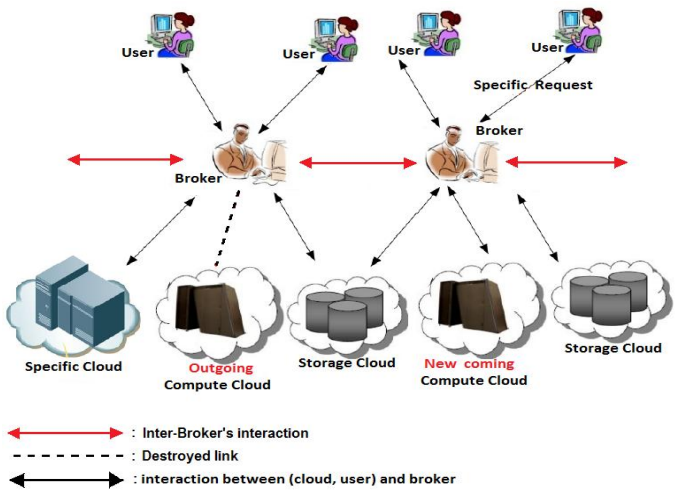

Fig.1 Flexible System Model

Three types of agents are considered in figure1: Consumer Agent, Resource Brokering Agent and Resource Provider Agent [2]. RBA will assume the complicate resource allocation task between providers and users in clouds federation. in [2] Haresh M V et. al. assume in their system model that The Broker Agent contains all information about all cloud providers of federated clouds. Even though due to dynamic feature in open federated clouds the Broker Agent can't have information about all the cloud providers constituting the federation. So each Broker Agent has his own contact's list of cloud providers which can be similar or different (more or less rich) from other contact's list neighbors belonging to Broker Agent society. The Broker Agent contains all information about cloud providers in his contact's list as location, prices, lowest cost of the resources and the provider which provides high quality of service in his own contact's list. The Broker Agent assigns a grade to the providers based on the feedback from the consumers to which it has done the allocations. All the time the Broker Agent will be checking the status of each of the cloud providers in his own list. The Broker Agent negotiates with Resource Provider Agents and if the requirement of a Consumer Agent is not fulfilled by a single provider, it initiates a negotiation with another Resource Provider Agent belonging to his contact's list.[2]

We notice in Flexible System Model interactions between RBA organization members (inter organization communications). This links permit request's migration between RBA neighbors, to delegate costumer requests in case where RBA, and even after checking all clouds provider contact's list, is incapable to satisfy the client resource demand and can't find the client request resource. It is called failure situation (i.e., resource still unavailable, price not suitable, specific provider leaves the federation, entry of new provider with a particular business or technological need which is still not well known by all RBA...).

For a problem modeling we will assume the same one in [2], in addition to:

.CA_id - Consumer Agent identification.

### A. RBA Self organization

In [5] an organization partitioning approach is proposed by adapting particles approach algorithm of Heiss et. al.[9] to role partitioning between agents system. We will adopt the concept of migration in organization partitioning approach as auto adaptive mechanism, then adapt the algorithm presented by Lahlouhi [5] to apply self organization in our RBA organization. We assume that links between RBA are fixed, a dominance relation is used to resolve conflict of choosing RBA destination or direction between neighbors. So the non satisfied client request will migrate from RBA source in the direction of the non dominate RBA neighbor in order to be fulfilled by a new RBA host. After the chosen of RBA direction is done, RBA source sent the non satisfied client request to RBA direction with some information linked to resources requests and other information linked to user identification : CA_id (Consumer Agent identification) which permit to RBA direction to identify the costumer. So due to failure situation RBA source delegate to RBA direction the task of satisfying his initial consumer. From now RBA direction will communicate directly with CA concerned and try to satisfy his demand. If even the new RBA direction can't satisfy the delegated request, the adaptive mechanism is engaged again and so on. The costumer can't know that his request migrate from initial broker to another, the adaptive mechanism is done with transparency manner. Major advantage that we can take in account no fixed number of various conflicting criteria in choosing RBA destination. Non limited number of parameters, which can be incompatible, can be considered in solving conflict (i.e., $f1(i,k)$ : RBA workload parameter, $f2(i,k)$ : RBA rate transfer time...etc.).

After the choice of RBA direction is done and before migration authorization of non satisfied costumer request, we have first to check some preventive constraints to preserve a coherent RBA organization. Also after migration, other corrective constraints for coherent reestablishment must be checked. That way we ensure at any moment that we have a coherent system evolution. For example as preventive constraints, we have to avoid to migrate a non satisfied costumer request to a RBA direction if its provider contact's list is empty or if we know from the beginning that this RBA has no cloud provider in his list liable to satisfy the migrated user request. As corrective constraint, after migration of a non satisfied costumer request we have to update workload parameter for both RBA source and direction.

| RBA Self Organization Segment Part (adaptive mechanism) |
|---|
| **5.1.6.** <u>if</u>  (nbre_mig < max_mig ) <u>then</u><br>-----------------------------------------------------------------<br>*// Multi-criteria functions evaluations*<br>**5.1.6.1.** **Get information from neighbors**<br>**5.1.6.2.** f(0) = {0,0,...,0}<br>**5.1.6.3.** <u>for</u> all k ϵ neighbors <u>do</u><br>   **5.1.6.3.1.** f(k) = {f1(k),f2(k) ... fn(k)}<br>   <u>end</u><br>-----------------------------------------------------------------<br>-----------------------------------------------------------------<br>*// Searching non dominated neighbor verifying the constraints*<br>**5.1.6.4.** <u>repeat</u><br>   **5.1.6.4.1.** direct = q   with not dom (f(q),f(k))<br>           k ≠ q ^ k ϵ neighbors<br>   **5.1.6.4.2.** Chosen ← verify_constraint(direct)<br>   **5.1.6.4.3.** Remove f(direct)<br>   <u>until chosen</u><br>-----------------------------------------------------------------<br>-----------------------------------------------------------------<br>*// If chosen neighbor is not the source S, migrate reestablish //coherence*<br>**5.1.6.5.** <u>if</u> ( direct ≠ Source) <u>then</u><br>   **5.1.6.5.1.** Send CFP<CA_id,Rc,est,dlt,pl,S> to direct<br>   **5.1.6.5.2.** Re_establish_coherence (direct)<br>   <u>end</u><br>-----------------------------------------------------------------<br><u>end</u> |

*B. Negotiation protocol*

In Fig.1 three protocols are used for negotiation

- between CA, RBA and RPA.
- between RBA organization members.

We adapt specially Algorithm 2 and Algorithm 1 in [2] to deal with a dynamic open federated clouds environment. Algorithm 2 is adapted by : permitting first inter RBA organization communications then integrating an adaptive behavior in case where RBA  is incapable to satisfy user request it is called failure situation.

Algorithm 1 is the protocol used for negotiation of

- CA with RBA,
- CA with RPA.

Algorithm 2 is the protocol used for negotiation of

- RBA with CA,
- RBA with RPA,
- and RBA with other RBA neighbors of RBA organization.

Algorithm 3 is the protocol used for negotiation of

- RPA with RBA,
- RPA with CA.

Notice that there is another Algorithm 4 for resource allocation. [2]

Consumer agent initialize the CFP message including his identification. CA send its CFP message to RBA with CA identification. RBA extract the resources from the consumer request. RBA update his current provider list of contacts (to take in account a probably new entrance or departure of providers). RBA initialize a temporary provider contact's list with his current provider contact's list and it searches each resource in his own temporary provider contact's list based on, maximum QoS and minimum unit cost to select the best provider. After RBA gets the resource list, it calculates the total cost of resource set Rc by  $\sum_{r \in Rc} q(Rc,r) c(r) p$  and send to Consumer Agent by PROPOSE message. If the cost is acceptable CA sends the ACCEPT PROPOSAL to RBA. If the cost is higher than expected cost CA sends the expected cost of resources by REJECT PROPOSAL.[2]

If RBA receive a REJECT PROPOSAL message from CA, RBA remove the best selected provider from his temporary provider contact's list and check the temporary contact's list, if it isn't empty RBA repeat the protocol from the start to select another best provider from the temporary contact's list. But in case where his temporary provider contact's list is empty the adaptation mechanism is engaged automatically by executing RBA Self Organization Segment part.[2]

RBA receive the ACCEPT PROPOSAL message from CA, then initializes a CFP message and sends to Resource Provider Agent. RPA receives the CFP message and extracts the resources and its total cost. Then calculate the expected cost of requested resources by equation(1) in [2] modeling problem. If the expected cost is less than or equal to CFP cost then accept the request and run the resource allocating algorithm for requested resource that is available at that time. If it is available for assignment, send PROPOSE message to RBA. If the expected cost is higher than requested cost then send the current demand/price ratio to RBA by REFUSE message. RBA update both temporary and current provider contact's list with this demand price ratio and repeat the same procedure.[2]

If RBA receives PROPOSE message from RPA then send the partial agreement to CA by PROPOSE message. If the agreement acceptable, CA send AGREE message to RBA and RBA send the CONFIRM message to RPA for agreement confirmation. If the agreement is not acceptable, CA send a refuse message to RBA so RBA remove the best selected provider from a his temporary provider contact's list, check the temporary contact's list, if it isn't empty RBA repeat the protocol from the start to select another best provider from the temporary contact's list. But in case where his temporary provider contact's list is empty the adaptation mechanism is engaged automatically by executing  RBA Self Organization Segment part.[2]

When RPA get the confirmation, it sends the final agreement to CA by CONFIRM message. And CA sends the acknowledge by INFORM message. After reaching the final agreement, the consumer agent begins the execution. After completion of the task, CA calculates the utility of the resources and sends it as feedback to RBA. RBA updates his own provider contact's list based on the received value.

In case where RBA execute RBA Self Organization Segment part then after finishing RBA update his provider contact's list by a feedback about the failure situation.[2]

---

**Algorithm 1: Consumer Agent Communication**

**begin**
 1. **initialize CA_id,Rc,est,dlt,pl,S** // *Include CA identification in*
  //   *initialization*
 2. **Send CFP < CA_id,Rc,est,dlt,pl,S> to RBA** // *Send CFP to RBA*
  //   *with CA_id*
 3. **Rceive PROPOSE (Cost of Rc c(Rc)) from RBA** // *Cost of Rc*
 4. **if cost acceptable then**
    4.1. **Send "ACCEPT_PROPOSAL" to RBA**
  **else**
    4.2. **Send "REJECT_PROPOSAL(cost-limit)" to RBA**
    4.3. **Goto step 3.**
  **end**
 5. **Rceive PROPOSE from RBA**
 6. **if confirmed then**
    6.1. **Send "AGREE" to RBA**
  **else**
    6.2. **Send "REFUSE" to RBA**
    6.3. **Goto step 5.**
  **end**
 7. **Receive "CONFIRM" from RPA**
 8. **Accept agreement and send "INFORM" to RPA**
 9. **Running task ...**
 10. **Calculate utility by eq(1)**
 11. **Send feedback "INFORM" to RBA**
**end**

---

**Algorithme 2: RBA Communication**

**begin**
 1. **Receive CFP < CA_id,Rc,est,dlt,pl,S> from CA or RBA**
 2. **Extract ressource from Rc**
 ---------------------------------------------------------------------------------
 3. **Update provider contact's list** // *permit provider's movement in*
  // *open federated clouds (entrance and departure)*
 ---------------------------------------------------------------------------------
 4. **Temporary_provider_list ← Provider_list**
 5. **if not(Temporary_provider_list is empty) then**
    5.1. **Select the best provider from Temporary_provider_list**
    5.2. **Calculate cost of each resource in Rc**
    5.3. **Send "PROPOSE(c(Rc))" to CA**
    5.4. **Receive ack from CA**
    5.5. **if "ACCEPT_PROPOSAL" then**
       5.5.1. **Send CFP < CA_id,Rc,est,dlt,pl> to RPA**
     **else**
       5.5.2. **Receive "REJECT_PROPOSAL(cost-limit)"**
       **from CA**
       5.5.3. **Remove best provider from**
        **Temporary_provider_list**
       5.5.4. **Goto step 5.**
     **end**
  **else**
  ---------------------------------------------------------------------------------
    5.6. **RBA Self Organization Segment part** // *Failure situation*
     // *adaptive machanism engaged*
  ---------------------------------------------------------------------------------
    5.7. **Goto step 11.**
  **end**
 6. **Receive response from RPA**
 7. **if PROPOSE then**
    7.1. **Send "PROPOSE" request to CA**
  **else**
    7.2. **Update cost resources in Temporary_provider_list and**
     **provider_list**
    7.3. **Goto step 5.2.**
  **end**
 8. **Receive ack from CA**
 9. **if "AGREE" then**
    9.1. **Send "CONFIRM" to RPA**
  **else**
    9.2. **Send "REFUSE" to RPA**
    9.3. **Remove best provider from Temporary_provider_list**
    9.4. **Goto step 5.**
  **end**
 10. **Receive "INFORM" from CA**
 11. **Update provider_list with feedback**
**end**

## C. Data flow diagram

In Fig.2 we illustrate negotiation between the three protocols. We adapt the data flow diagram in [2] first by updating the provider contact's list before start searching the best provider in order to deal with open federated clouds. After that in case of failure situation (providerlist is empty) we execute RBA Self Organization mechanism, a feedback about failure is sent to providers. CA request will migrate to the new RBA direction and the same process will be repeated again from the begining untill satisfaction. The custmer have no need to know who performe his request, so he don't feel that his request is migrating from a broker to another. In final the user's request will be done and self organization mechanism is executed discreetly.

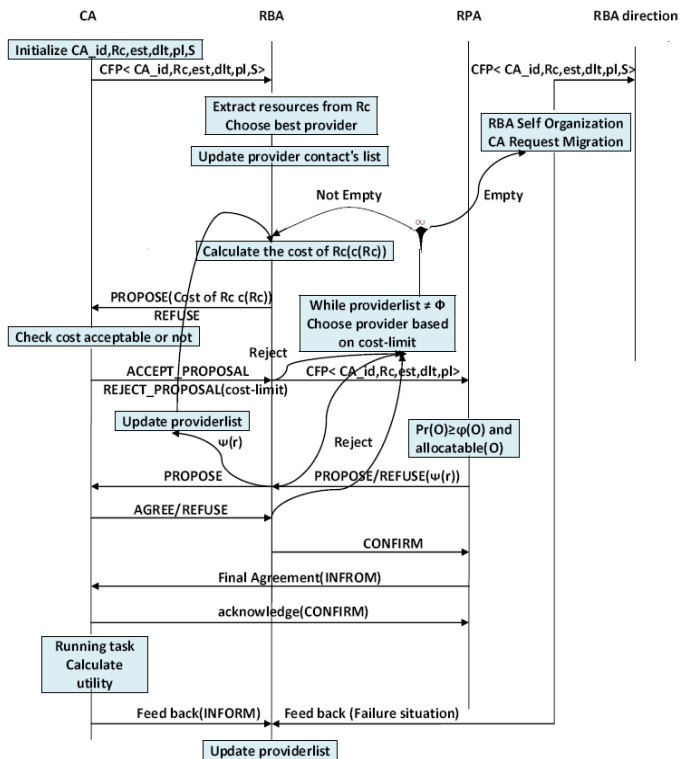

Fig.2 Data flow diagram

## IV. CONCLUSION AND FUTURE WORKS

Actually business is a very dynamic environment, new cloud providers can enter to the market and offer their new services, while other cloud providers can leave it. We consider in this paper an open clouds federation environment which is in constant evolution, new cloud operators can join the federation or leave this one. We tried to present a robust approach to deal with changes of open federated cloud computing environment by proposing an auto-adaptive method. Starting by adapting two already existing approaches with open federated clouds environment before combining between them. The first one is organizations optimization approach for self organization in broker agent organization, second one is Multi Agent Resource Allocation approach on Federated Clouds. Our approach is auto adaptive and multi criterion which can take in account various parameters (i.e. computing load balance of mediator agent, geographical distance (network delay) between costumer and provider...).

For the future we are looking at a concrete implementation to test the robustness of the system using JADE[10], but also observe emergent behavior of broker mediator organization. Then providing new features in open federated clouds accomplishment.